\documentclass[twocolumn,prb,showpacs,preprintnumbers,amsmath,amssymb]{revtex4}
\usepackage{graphicx}% Include figure files
\usepackage{dcolumn}% Align table columns on decimal point
\usepackage{bm}% bold math
\usepackage{epsfig}
\begin{document}

%\preprint{APS/123-QED}

\title{On the mechanism of the highly viscous flow}% Force line breaks with \\

\author{U. Buchenau}
\email{buchenau-juelich@t-online.de}
\affiliation{%
Institut f\"ur Festk\"orperforschung, Forschungszentrum J\"ulich\\
Postfach 1913, D--52425 J\"ulich, Federal Republic of Germany}%
\date{May 18, 2011}% It is always \today, today,

\begin{abstract}
The asymmetry model for the highly viscous flow postulates thermally activated jumps from a practically undistorted ground state to strongly distorted, but stable structures, with a pronounced Eshelby backstress from the distorted surroundings. The viscosity is ascribed to those stable distorted structures which do not jump back, but relax by the relaxation of the surrounding viscoelastic matrix. It is shown that this mechanism implies a description in terms of the shear compliance, with a viscosity which can be calculated from the cutoff of the retardation spectrum. Consistency requires that this cutoff lies close to the Maxwell time. The improved asymmetry model compares well with experiment.  
\end{abstract}

\pacs{64.70.Pf, 77.22.Gm}% PACS, the Physics and Astronomy

\maketitle

\section{Introduction}

There are many theories and models of the flow process \cite{heuerr,adam,ngai,gotz,shoving,diezemann,pick1,wolynes,facil,paluch,granato,bouchaud,sokolov,berthier,pick2,johnson,nemilov,dyre,johari,langer}, but two central points are still not clear. The first is the detailed development of the viscosity from the thermally activated jumps between inherent states \cite{goldstein,stillinger} in the highly viscous liquid. The second is the role of the elastic interaction between different jumps. 

The  phenomenological asymmetry model \cite{asymm} focuses precisely on these two central points. It is able to explain the asymmetry observed in aging experiments at the secondary relaxation peak \cite{olsen}. In these experiments, one gets a sudden small rise of the peak height after a small temperature rise even before the energy landscape adapts itself to the new temperature. This rise is only understandable if there is an average energy difference of about 4 $k_BT$ between the structural minima of the secondary relaxation.

The asymmetry model ascribes the asymmetry to a structural rearrangement of an inner core (a "cooperatively rearranging region" \cite{adam}), consisting of $N$ atoms or molecules, with an $N$ of the order of hundred. The structural rearrangement leads to a change of shape, which in turn leads to an elastic distortion, both of the core and of the surrounding matrix, an elasticity theory problem solved by Eshelby \cite{eshelby}. The Eshelby concept, first introduced into the field to explain the plastic deformation of metallic glasses \cite{harmon}and closely related to the concept \cite{langer} of shear transformation zones, is a good starting point for a quantitative description of the flow process, not only because it provides a quantitative relation between the local relaxational jump and its coupling constant to an external shear, but also because it allows for a simple mean-field treatment of the interaction between different relaxing entities.

According to the asymmetry model, the flow proceeds by a jump from the unstrained ground state over a relatively high barrier into a distorted state. If the barrier is high, the inner domain remains for a long time in the distorted state. If this time is long enough, the inner core can relax by relaxation processes in the neighborhood, to become a new unstrained ground state.

If this is in fact the appropriate physical basis, the calculation of the shear response in the former treatment \cite{asymm} can be considerably improved by relating the viscosity to the cutoff of the retardation spectrum. This is the aim of the present paper.

The following section II begins with a reformulation of the asymmetry model in terms of the shear compliance, deriving expressions for the viscosity and the retardation spectrum. Section III describes fits to experimental data from the literature. Section IV discusses the success and the limitations of the treatment, section V contains the summary. 

\section{The compliance formulation of the asymmetry model}

One has two equivalent textbook descriptions of the shear spectrum of a liquid \cite{ferry}, a relaxation spectrum $H(\tau)$ for the description of the complex shear modulus $G(\omega)$ ($\tau$ relaxation time)
\begin{equation}\label{gom}
G(\omega)=\int_{-\infty}^\infty \frac{H(\tau)i\omega\tau}{1+i\omega\tau}d\ \ln\tau
\end{equation}
and a retardation spectrum $L(\tau)$ for the description of the complex shear compliance
\begin{equation}\label{jom}
J(\omega)=\frac{1}{G_\infty}+\int_{-\infty}^\infty \frac{L(\tau)}{1+i\omega\tau}d\ \ln\tau-\frac{i}{\omega\eta}.
\end{equation}

Here $G_\infty$ is the infinite frequency shear modulus and $\eta$ is the viscosity. A third material constant hidden in this equation is the recoverable compliance $J_e^0$, the elastic compliance plus the integral over the retardation processes
\begin{equation}
	J_e^0=\frac{1}{G_\infty}+\int_{-\infty}^\infty L(\tau)d\ \ln\tau.
\end{equation}

Eq. (\ref{jom}) makes a separation of two independent contributions to the compliance, the retardation spectrum and the viscosity. The retardation spectrum is due to back-jumps into the initial inherent state, the viscosity is due to no-return processes. The retardation description separates these two influences, the relaxation description does not. In the relaxation description of eq. (1), $\eta$ is given by
\begin{equation}\label{etagom}
\eta=\int_{-\infty}^\infty \tau H(\tau) d\ \ln\tau
\end{equation}
and is an inherent property of the spectrum.

In principle, one could also separate $H(\tau)$ into a flow and a relaxation component, but the truncation would be far less obvious than in the $L(\tau)$ case. A formally similar problem is encountered in the field of dielectric relaxation, where the dielectric modulus could also in principle include a zero frequency component. But the usual way is to add an independent conductivity term to the dielectric susceptibility.
  
Since the asymmetry model \cite{asymm} predicts an independent flow process (though mediated by the retardation processes), it should be formulated in terms of the compliance description. This was not done in the original derivation \cite{asymm}, but will be done here. As a further improvement, the Poisson ratio \cite{sokolov,johari} is explicitly taken into account.

The asymmetry model \cite{asymm} considers an embedded core volume of $N$ atoms or molecules (atomic or molecular volume $v_a$) which can undergo structural rearrangements. The shape difference of the inner volume $Nv_a$ is either a shear $e_i$ (the shear angle between the two stable structures of the inner volume in radian) or a volume change $\delta v$ or a mixture of both. For a shear $e_i$ and an infinitely hard surrounding medium, the asymmetry (the energy difference between final and initial state) $\Delta$ would be $G_\infty Nv_ae_i^2/2$, where $G_\infty$ is the (infinite frequency) shear modulus. But we consider a surrounding medium with the same elastic constants as the inner part. In this case, the medium distorts as well and one gets a lowering of the elastic energy by about a factor of two, slightly depending on the Poisson ratio $\nu$ of the substance \cite{eshelby}
\begin{equation}\label{d}
    \Delta=\gamma_pG_\infty Nv_a\frac{e_i^2}{2},
\end{equation}
where for a spherical inner volume the Eshelby-Poisson reduction factor $\gamma_p$ is given by
\begin{equation}\label{gammap}
	\gamma_p=\frac{7-5\nu}{15(1-\nu)}
\end{equation}
and the coupling to an external shear $e$ in the direction of $e_i$ is given by
\begin{equation}\label{dde}
    \frac{\partial\Delta}{\partial e}=\gamma_pG_\infty Nv_ae_i.
\end{equation}
This implies that the influence of the Poisson ratio can be taken into account by replacing $G_\infty v_a$ in the equations of the earlier paper \cite{asymm} by $2\gamma_pG_\infty v_a$.

A given asymmetry $\Delta$ defines the surface of a sphere in the six-dimensional distortion space. For a constant density of stable structures in distortion space, one gets a number of states $n(\Delta)d\Delta$ between $\Delta$ and $\Delta+d\Delta$
\begin{equation}
    n(\Delta)=\frac{c_N}{N^3}\frac{\Delta^2}{(2\gamma_pG_\infty v_a)^3},
\end{equation}
where $c_N$ is a dimensionless and temperature-independent number. The partition function
\begin{equation}\label{zdef}
    Z=1+\int_0^\infty\frac{c_N}{(2\gamma_pG_\infty Nv_a)^3}\Delta^2{\rm e}^\frac{-\Delta}{k_BT}d\Delta
\end{equation}
is easily evaluated. With the definition of the characteristic multi-minimum parameter $f_N$
\begin{equation}\label{fn}
    f_N=\frac{c_N}{N^3}\left(\frac{k_BT}{2\gamma_pG_\infty v_a}\right)^3,
\end{equation}
the partition function reads
\begin{equation}\label{z}
    Z=1+2f_N,
\end{equation}
where the 1 stands for the ground state and the $2f_N$ for the distorted excited states.

The second order term $Z_e$ results from the integral 
\begin{equation}\label{ze2}
    Z_e=\frac{f_Ne^2\gamma_pG_\infty Nv_a}{6k_BT}\int_0^\infty p_{\delta G}(\Delta)d\Delta,
\end{equation}
where the function $p_{\delta G}(\Delta)$ is given by
\begin{equation}\label{deltaG}
	p_{\delta G}(\Delta)=\frac{\Delta^3}{(k_BT)^4}{\rm e}^\frac{-\Delta}{k_BT}.
\end{equation}

Since this function is the same as in the treatment without the Eshelby-Poisson factor $\gamma_p$, the conclusion of an average asymmetry of $4k_BT$ remains unchanged.

To second order, the partition function with an applied strain $e$ is
\begin{equation}\label{ze}
    Z=1+2f_N+\frac{f_N\gamma_pG_\infty Nv_a}{k_BT}e^2
\end{equation}

With eq. (\ref{ze}), one can calculate the contribution $\delta G_N$ of the volume $Nv_a$ to the reduction of the shear modulus, using the free energy $F=-k_BT\ln Z$ and $\delta G_N=(1/Nv_a)\partial^2F/\partial e^2$
\begin{equation}\label{dgn}
    \frac{\delta G_N}{G_\infty}=-\frac{2\gamma_pf_N}{1+2f_N}.
\end{equation}
This is the reduction of the shear modulus {\it without} interaction between different structural rearrangements.

Let us next consider the influence of the dynamics, characterized by the variable $v=\ln\tau_v$, where $\tau_v$ is the relaxation time. Let us denote by $f_0(v)$ the barrier density of structural rearrangements of the $N$ molecules or atoms, in the sense that $-\delta G_N/G_\infty$ from these relaxations (without interaction) between $v$ and $v+dv$ is given by $f_0(v)dv$.

With this definition, it turns out to be surprisingly easy to take the elastic interaction between different relaxation centers into account. At the time $\tau_v$, the shear modulus is reduced by all the relaxation centers with shorter relaxation times than $\tau_v$ to the value $G_v$. For a given relaxation, this means that its asymmetry $\Delta$ is reduced from its initial value by the factor $G_v/G_\infty$ (assuming that the inner domain stress energy decreases by the same amount as the stress energy outside). Since the total effect for all relaxations between $v$ and $v+dv$ is an integral over the full distortion space, one gets again the integral over the function $p_{\delta G}(\Delta)$ of eq. (\ref{ze2}). If all $\Delta$-values decrease by $G_v/G_\infty$, the integral increases by $(G_\infty/G_v)^4$.

On the other hand, the coupling constant to an external strain $e$ also decreases by the factor $G_v/G_\infty$. Since the coupling constant enters with its square into the decrease of the shear modulus, the barrier density $f(V)$ {\it with} interaction must be related to the barrier density $f_0(V)$ {\it without} interaction by
\begin{equation}\label{inter}
	f(v)=\frac{G_\infty^2}{G_v^2}f_0(v).
\end{equation}
The solution to this equation has been given earlier \cite{philmag}. For a given $f_0(v)$, $f(v)$ reads
\begin{equation}\label{fv}
	f(v)=\frac{f_0(v)}{\left[1-3\int_0^v f_0(v')dv'\right]^{2/3}},
\end{equation}
a solution which brings the shear modulus down to zero if $f_0(v)$ integrates to 1/3 (the 1/3-rule).

But the essential point of the present paper, the one that was not understood in the previous attempts \cite{asymm,philmag}, is that the shear modulus does {\it not} go down to zero; it goes down to $1/J_e^0$ and stays there, while the independent viscosity processes do the rest.

This being so, one needs to formulate the problem in terms of the retardation spectrum $L(\tau)$ and the viscosity $\eta$. Let us do that.

Since the relaxation centers do not couple to the stress, but to the strain (see eq. (\ref{dde})), one has to formulate the description of their effect in terms of a relaxation spectrum like eq. (\ref{gom}), but with a finite rest modulus $1/J_e^0$ at the frequency zero
\begin{equation}\label{gom1}
G_r(\omega)=\frac{1}{J_e^0}+G_\infty\int_0^\infty \frac{f(v)i\omega\tau_v}{1+i\omega\tau_v}dv.
\end{equation}
$G_r(\omega)$ has to be inverted to obtain the shear retardation spectrum. Equation (\ref{gom1}) implies that the integral over $f(v)$ is limited to $1-1/G_\infty J_e^0$. Translating this into a limitation of the integral over $f_0(v)$
\begin{equation}\label{rj0e}
	\int_0^\infty f_0(v)dv=\frac{1}{3}-\frac{1}{(GJ_e^0)^3}.
\end{equation}
Obviously, the function $f_0(v)$ must have a cutoff at a critical $v_c$, with a critical relaxation time $\ln\tau_c=v_c$.

Denote the inherent relaxation density (the one without cutoff) by $f_i(v)$. Let these inherent relaxations decay with the time constant $\tau_c$. Then the cutoff function for $f_i(v)$ is given by $\tau_c/(\tau_v+\tau_c)$. The function $f_0(v)$ must result from the multiplication of the inherent relaxation density $f_i(v)$ with this cutoff function squared
\begin{equation}\label{fi}
	f_0(v)=\frac{\tau_c^2}{(\tau_v+\tau_c)^2}f_i(v)
\end{equation}
because only a fraction $\tau_c/(\tau_v+\tau_c)$ of $f_i(v)\tau_c/(\tau_v+\tau_c)$ returns to contribute to the retardation. The rest, a fraction $\tau_v/(\tau_v+\tau_c)$ of $f_i(v)\tau_c/(\tau_v+\tau_c)$, decays via the viscosity escape. Its contribution is only sizable if $\tau_v\approx\tau_c$. This opens up a way to calculate the viscosity.

For simplicity, the asymmetry energy $\Delta$ as well as all other energies will be counted in units of $k_BT$ in the following considerations.

To get the viscosity, assume a stationary flow with an applied shear stress $\sigma$. The "back-lag" $e_b$ of the equilibrium strain is $\sigma J_e^0$.  Take a relaxation center with an asymmetry $\Delta$ which happens to have its inner structural shear displacement $e_i$ in the direction of the applied shear stress. In the stationary flow, the population $p$ of the distorted state is increased by the factor
\begin{equation}\label{pop}
	1+e_b\frac{\partial\Delta}{\partial e}=1+\sigma J_e^0\sqrt{4\Delta\gamma_pG_\infty Nv_a},
\end{equation}
where equs. (\ref{d}) and (\ref{dde}) have been used.

If the surroundings of the distorted state with energy $\Delta$ relax, it becomes a new unstrained ground state. In the transition, the volume $Nv_a$ flows by the shear angle
\begin{equation}
	e_i=\sqrt{\frac{2\Delta}{\gamma_pG_\infty Nv_a}}.
\end{equation}
One then gets a contribution to the flow
\begin{equation}\label{ddote}
	\delta \dot{e}=2\Delta\sigma J_e^0\frac{p}{\tau_c}.
\end{equation}
Note that the number $N$, the Eshelby-Poisson factor and $G_\infty$ disappear in the product. If one has to replace $G_\infty$ by $G_v$, the equation remains valid.

One gets the same contribution from a relaxation center oriented in the opposite direction, because both the population change and the flow direction change sign. For the generally oriented defect, one gets a factor of $1/6$ from the six components of the distortion, one compression and five shears.

Integrating over $\Delta$, one gets a factor of 3 (a factor of 6 from the integral over $p_{\delta G}$ and a factor of 1/2 from the normalization). Summing up over all relaxation centers
\begin{equation}
	\dot{e}=\frac{\sigma J_e^0}{\gamma_p\tau_c}\int_0^\infty \frac{\tau_v\tau_c}{(\tau_c+\tau_v)^2}f_i(v)dv,
\end{equation}
where the $\gamma_p$ in the prefactor results from the weight of the excited states in the partition function, equs. (\ref{ze}) and (\ref{dgn}).

Since $\dot{e}=\sigma/\eta$, this supplies an equation for the Maxwell time $\tau_M=\eta/G_\infty$
\begin{equation}\label{taum}
	\frac{1}{\tau_M}=\frac{G_\infty J _e^0}{\gamma_p\tau_c}\int_0^\infty\frac{\tau_v}{\tau_c}f_0(v)dv.
\end{equation}

The same integral supplies the lifetime $\tau_{crr}$ of the cooperatively rearranging region, because a fraction $\tau_v/\tau_c$ of $f_0(v)$ decays with the rate $1/\tau_c$. Therefore
\begin{equation}\label{tcrr}
	\tau_{crr}=G_\infty J_e^0\tau_M.
\end{equation}
This lifetime is considerably longer than $\tau_c$, because only the excited states decay and even there, the decay competes with the back-jump probability.

The weight factor $f_{gs}$ of the undistorted ground state in the canonical ensemble can be calculated from equs. (\ref{z}) and (\ref{dgn}). One finds
\begin{equation}\label{fgs}
	f_{gs}=1-\frac{1-1/(G_\infty J _e^0)^3}{3\gamma_p},
\end{equation}
a value close to 1/3.

In order to achieve consistency, the decay rate $1/\tau_c$ should be ascribed to the viscous flow outside the inner core. As pointed out by Eshelby \cite{eshelby}, one has a fraction $\gamma_p$ of the total elastic energy in the core and a fraction $1-\gamma_p$ outside. This implies a decay rate for the asymmetry $\Delta$ by the viscosity outside
\begin{equation}
	\frac{\dot{\Delta}}{\Delta}=\frac{1-\gamma_p}{\tau_M}
\end{equation}
The distortion lifts the energy of the distorted state by $\Delta$. The energy increases quadratically with the distance from the ground state in distortion space. Therefore it lifts the energy at the barrier between the two states by $\Delta/4$, and the relative change of $\tau_v$ is given by
\begin{equation}
	\frac{\dot{\tau_v}}{\tau_v}=\frac{3\Delta}{4}\frac{1-\gamma_p}{\tau_M}.
\end{equation}

$\tau_c$ was defined as the relaxation time at which an equal number of inherent states goes into both channels, the viscosity and the retardation. This implies the condition $\dot{\tau_v}=1$ at the decay time constant $\tau_c$, which should be obeyed at the average value $\Delta=4$, so the consistency condition reads
\begin{equation}\label{cons}
	\frac{\tau_M}{\tau_c}=3(1-\gamma_p).
\end{equation}
This means $\tau_m$ is factor of 3/2 larger than $\tau_c$ if the Eshelby-Poisson factor is exactly 1/2. In any case, $\tau_c$ lies close to the Maxwell time $\tau_M$.

%%%%%%%%%%%%%%%%%%%%% begin figure %%%%%%%%%%%%%%%%%%%%%%%%%%%%%%%%%%%%%
\begin{figure}[b]
\hspace{-0cm} \vspace{0cm} \epsfig{file=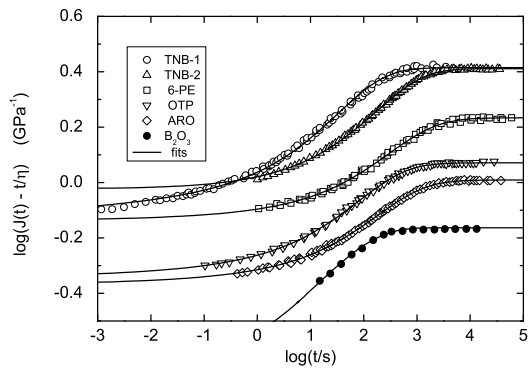,width=7cm,angle=0} \vspace{0cm}\caption{Asymmetry model fits of the recoverable compliance of six glass formers at the glass transition \cite{plazek-magill,plazek-bero,tnb,plazek-bo}. Notation see Table I (ARO is Aroclor \# 1248 and OTP is orthoterphenyl).} 
\end{figure}
%%%%%%%%%%%%%%%%%%%%% end figure %%%%%%%%%%%%%%%%%%%%%%%%%%%%%%%%%%%%%%%

\section{Comparison to experiment}

Even with the equations derived in the preceding section, one is left with five unavoidable independent parameters with a well-defined physical meaning: the infinite frequency shear modulus $G_\infty$, the viscosity $\eta$, the recoverable shear compliance $J_e^0$, the Eshelby-Poisson factor $\gamma_p$ and a stretching parameter $\gamma$ (the analog of the Kohlrausch $\beta$).

For a problem-adapted fitting procedure, one chooses $G_\infty$, $\gamma_p$, $G_\infty J_e^0$ and $v_c=\ln\tau_c$ as four of the five independent parameters. $v_c$ stands for the viscosity, because one can calculate $\eta$ via the Maxwell time from eq. (\ref{cons}).

The stretching parameter $\gamma$ is introduced into the equation for $f_0(v)$
{\begin{equation}\label{f0v}
	f_0(v)=a\exp(\gamma(v-v_c)+\gamma_2(v-v_c)^2)\frac{\tau_c^2}{(\tau_c+\tau_v)^2}
\end{equation}
where the factor $a$ and the curvature $\gamma_2$ are not free parameters, but fixed by equs. (\ref{rj0e}) and (\ref{taum}). If $\gamma_2$ is negative, the function is a gaussian with a cutoff at $v_c=\ln\tau_c$.

If one has a pronounced secondary peak, one needs the position and the width of this peak as additional parameters. In this case, the height of the peak takes the role of the curvature parameter $\gamma_2$, which is set to zero. 

The most stringent test for the model is supplied by the compliance measurements of Plazek and coworkers \cite{plazek-magill,plazek-bero,tnb,plazek-bo}, because these measurements separate retardation and viscosity explicitly. Fig. 1 shows the fit of the recoverable compliance data (the recoverable compliance $J(t)-t/\eta$ is only due to the retardation) for five molecular glass formers and for boron trioxyde. In boron trioxyde, the shear modulus $G_\infty$ of the fit was taken from Brillouin \cite{grimsditch} and ultrasonic data \cite{macedo}. As it turns out, the recoverable compliance data do not fix the Eshelby-Poisson-factor $\gamma_p$ very accurately, so one can still choose it rather freely. The choice is nailed down by the measured viscosity, which together with the recoverable compliance fixes the Eshelby-Poisson parameter with an accuracy of about 1 to 2 \%. This joint fit requires a non-zero curvature $\gamma_2$, which is sometimes positive and sometimes negative.

%%%%%%%%%%%%%%%%%%%%% begin figure %%%%%%%%%%%%%%%%%%%%%%%%%%%%%%%%%%%%%
\begin{figure}[b]
\hspace{-0cm} \vspace{0cm} \epsfig{file=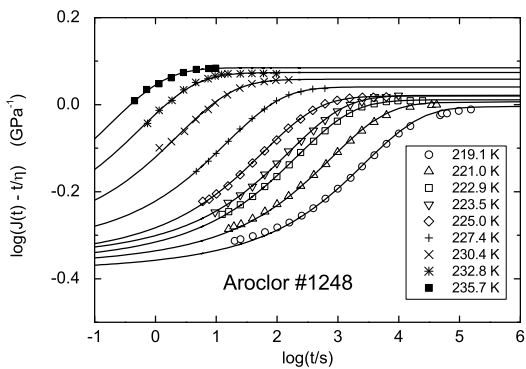,width=7cm,angle=0} \vspace{0cm}\caption{Asymmetry model fits of the recoverable compliance of Aroclor \# 1248 at different temperatures \cite{plazek-bero}.}
\end{figure}
%%%%%%%%%%%%%%%%%%%%% end figure %%%%%%%%%%%%%%%%%%%%%%%%%%%%%%%%%%%%%%%

Looking at data from the same substance at different temperatures, one can check whether the fitted Eshelby-Poisson parameter is temperature dependent. In the case of Aroclor \cite{plazek-bero} in Fig. 2, one finds no visible temperature dependence even within the small experimental error.

The error bars are slightly larger (about 5 to 10 \%), if one fits measurements of $G(\omega)$. Fig. 3 (a) shows propylene glycol \cite{maggi} and Fig. 3 (b) tripropylene glycol \cite{niss} as examples for substances without and with visible secondary relaxation peak. In the case of tripropylene glycol one can take position and width from dielectric measurements \cite{niss} and replace the parameter $\gamma_2$ by the height of the peak, so one has again five parameters. 

The fit of propylene glycol in Fig. 3 (a) provides a negative $\gamma_2= -0.012$. Setting this value constant at all measured temperatures, one gets a practically temperature-independent $\gamma=0.10$. This implies that one sees a gaussian in $f_0(v)$ with its maximum at $v_c+4.1$ at all temperatures. In other words, the maximum of the barrier density $f_0(v)$ shifts with temperature in such a way that it keeps the constant relaxation time ratio $\tau_{max}/\tau_c=64.5$ at all temperatures, where $\tau_{max}$ is the relaxation time at the maximum. The same is found for the data \cite{niss} of triphenylethylene, only with the slightly larger ratio $\tau_{max}/\tau_c=181$.

This kind of peak shift is expected in the elastic models \cite{nemilov,dyre}, which postulate a proportionality between the barrier heights and the infinite frequency shear modulus $G_\infty$. What does not fit is the temperature-independent curvature $\gamma_2$, which should increase with $T^2/G_\infty^2$ with increasing temperature. Such a decrease can be excluded within experimental error in both cases; the fits rather indicate a broadening with increasing temperature. The finding is relevant for the question whether the temperature dependence of $G_\infty$ can indeed account for the full fragility \cite{bu2009}. 

The model gives excellent fits. In fact, the propylene glycol fit in Fig. 3 (a) is by no means the best, but belongs to the poorer ones of all those done to obtain the data in Table I.

%%%%%%%%%%%%%%%%%%%%% begin figure %%%%%%%%%%%%%%%%%%%%%%%%%%%%%%%%%%%%%
\begin{figure}[b]
\hspace{-0cm} \vspace{0cm} \epsfig{file=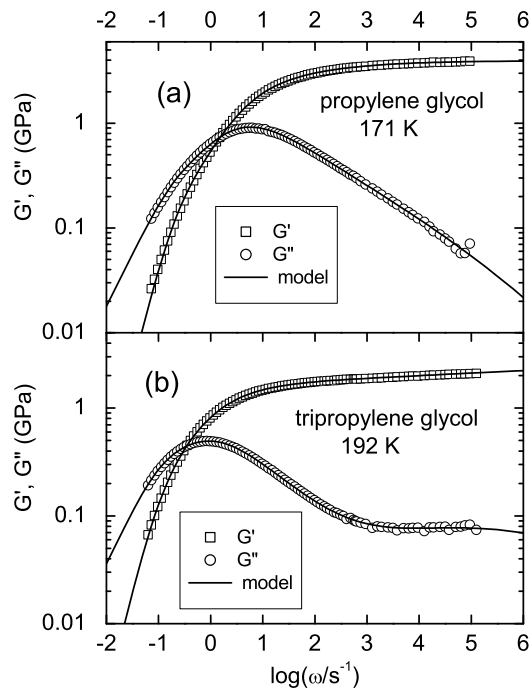,width=7cm,angle=0} \vspace{0cm}\caption{Fits of $G(\omega)$ in terms of the new version of the asymmetry model in a) propylene glycol b) tripropylene glycol.} 
\end{figure}
%%%%%%%%%%%%%%%%%%%%% end figure %%%%%%%%%%%%%%%%%%%%%%%%%%%%%%%%%%%%%%%

Table I compiles the fitted values of $G_\infty J _e^0$ and the Eshelby-Poisson factor $\gamma_p$. In those cases where measurements at different temperatures were available, the Aroclor result of a temperature-independent $\gamma_p$ was confirmed within experimental accuracy. The recoverable compliance increased with increasing temperature in the two cases dibutylphthalate and tripropylene glycol, both substances with a secondary relaxation peak, and decreased with increasing temperature in glycerol and meta-toluidine.

\begin{table}[htbp]
	\centering
		\begin{tabular}{|c|c|c|}
\hline
substance                                                        & $\gamma_p$         & $G_\infty J _e^0$  \\
\hline   
                                                                 &                    &           \\
\hline                                                                                           
silica \cite{mills}                                              & 0.48$\pm$0.1       &    2.0    \\
boron trioxyde \cite{plazek-bo}                                  & 0.88$\pm$0.01      &    3.6    \\
Zr$_{65}$Al$_{7.5}$Cu$_{17.5}$Ni$_{10}$ \cite{samwer}            & 0.50$\pm$0.03      &    2.5    \\
Pd$_{40}$Ni$_{40}$P$_{20}$ \cite{samwer}                         & 0.50$\pm$0.04      &    3.6    \\
orthoterphenyl \cite{plazek-bero}                                & 0.57$\pm$0.01      &    2.61   \\
Aroclor \# 1248 \cite{plazek-bero}                               & 0.59$\pm$0.01      &    2.36   \\
TNB-1 \cite{plazek-magill}                                       & 0.55$\pm$0.01      &    2.7    \\
TNB-2 \cite{tnb}                                                 & 0.56$\pm$0.01      &    3.5    \\
6-PE \cite{plazek-bero}                                          & 0.60$\pm$0.01      &    2.35   \\
PPE \cite{niss}                                                  & 0.59$\pm$0.05      &    2.35   \\
glycerol \cite{donth,jeong}                                      & 0.57$\pm$0.05      &  6.5-3.5  \\
propylene carbonate \cite{donth2}                                & 0.51$\pm$0.1       &    3.1    \\
DGEBA \cite{donth2}                                              & 0.56$\pm$0.05      &    3.0    \\
m-toluidine \cite{maggi}                                         & 0.54$\pm$0.05      &  3.5-2.8  \\
propylene glycol \cite{maggi}                                    & 0.56$\pm$0.05      &    5.7    \\
diethyl phthalate \cite{maggi}                                   & 0.54$\pm$0.05      &    3.1    \\
dibuthyl phthalate \cite{maggi}                                  & 0.54$\pm$0.05      &  2.7-3.3  \\
DC704 \cite{niss}                                                & 0.55$\pm$0.05      &    2.5    \\
triphenylethylene \cite{niss}                                    & 0.58$\pm$0.05      &    2.6    \\
tripropylene glycol \cite{niss}                                  & 0.52$\pm$0.05      &  3.1-4.3  \\
\hline		
		\end{tabular}
	\caption{Fitted Eshelby-Poisson factors and recoverable compliances for twenty glass formers. Aroclor \# 1248 is a chlorinated biphenyl of the company Monsanto; TNB-1 is 1,3-bis(1-naphtyl)-5-(2-naphtyl)benzene; TNB-2 is 1,3,5-tris(1-naphtyl)benzene; 6-PE is bis(m-(m-phenoxy phenoxy)phenyl)ether, a chain of six phenyl rings joined by oxygens; PPE, polyphenylene, is a commercial diffusion pump oil which is chemically very close to 6-PE; DGEBA is an epoxy resin, diglycidyl ether of bisphenol-A; DC704 is a silicon oil for diffusion pumps. The references denote the sources of the experimental data.}
	\label{tab:Comp}
\end{table}

The values of $\gamma_p$ of Table I look very reasonable. The value 0.57 found in orthoterphenyl is slightly larger than the one calculated from eq. (\ref{gammap}). The longitudinal and transverse sound velocities of orthoterphenyl at its glass transition temperature are \cite{toelle} $v_l=2.94$ km/s and $v_t=1.37$ km/s, respectively. The Poisson ratio $\nu = ((v_l/v_t)^2-2)((v_l/v_t)^2-1)/2$ is 0.361, so $\gamma_p=0.542$. One expects (and finds) a similar value for all the other molecular glass formers in Table I.

The value 0.48 of silica fits its low Poisson ratio of 0.12, while the value 0.88 for B$_2$O$_3$ is clearly too large to be explained by any possible Poisson ratio in eq. (\ref{gammap}). But there is a loophole for escape: The Eshelby-Poisson factor depends also on the shape of the core region. The point is discussed extensively by Eshelby \cite{eshelby}. To find such a large $\gamma_p$ as in B$_2$O$_3$, one needs two parallel extended planes shearing against each other in the core region. This could indeed happen in B$_2$O$_3$, which tends to form covalent planes. Some of the comparatively large $\gamma_p$-value of OTP and other aromatic compounds in Table I could be due to the same reason. In the same context, the value 0.5 for the two metallic glasses is interesting (from their Poisson ratio, they resemble the molecular glass formers). 0.5 is the exact value \cite{eshelby} for a needle lying either along x or y in an $e_{xy}$-shear (in this case, the needle is neither shortened nor lengthened by the shear). The finding agrees with the suggestion \cite{sam} of a flow in metallic glasses due to the strings seen first in numerical work on relaxation in glasses \cite{schober} and later in numerical work on the flow process in liquids \cite{glotzer}. 
    
\section{Discussion}

A first question is whether one can indeed do statistics with the small number of inherent states of a "cooperatively rearranging region" \cite{adam}, a relatively small volume of about hundred atoms or molecules. A recently developed four-point correlation method \cite{berthier} allows to estimate the size of this region from the temperature dependence of, say, the dielectric susceptibility \cite{cecile,roland,maggi2}. The average number of inherent states in such a cooperatively rearranging region can be calculated from the asymmetry model \cite{asymm}, because the $f_N$ for just two inherent states with asymmetry $4k_BT$ in the core is 0.012. To come to the measured values of $G_\infty J _e^0$ of 2 to 3, one needs $f_N\approx 0.7..0.9$, so there are sixty to eighty inherent states in a single cooperatively rearranging region. The size of the region increases strongly with decreasing temperature \cite{cecile,roland,maggi2}, but the number of inherent states remains fixed by the recoverable compliance $G_\infty J _e^0$, which according to the results in Table I tends to stay constant. The dramatic increase of the volume needed for a constant number of inherent states with decreasing temperature is consistent with the well-known dramatic entropy loss which leads to the extrapolated Kauzmann catastrophe. The ensemble of sixty to eighty states, though small, is large enough to make statistical considerations and to calculate a canonical partition function.

A further possible objection is that there is no such clear separation between outside and inside as assumed by the model; the next jump can occur partly inside and partly outside of the core region of the former jump. But the definition of the core region requires the unstrained ground state, which bundles the sixty to eighty jumps together (this is also the answer to another possible question, namely why there is no broad distribution of $\tau_c$-values).

The traditional theoretical procedure to calculate the shear response \cite{pick1,pick2} makes use of the Zwanzig-Mori formalism \cite{zwanzig}, considering appropriate projection operators which project the interesting variable out of the full set of bath variables and calculating their development in time via the Liouville operator. The appropriate variable in the shear case is the shear strain. Therefore the bath should be described in terms of a sum of relaxators reacting to the applied shear strain, equivalent to the $f(v)$ of the present paper (the Eshelby concept \cite{eshelby} leads to the same conclusion). The viscosity and the recoverable compliance are obtained from the first and second moments of this relaxation spectrum in relaxation time.

The analysis of the present work reveals the danger of the traditional approach. It induces one to believe that the relaxation density which describes the shear response reflects the bath processes via their coupling to the shear strain. According to the present analysis, this is only true in the initial stage, but is the wrong interpretation of the final loss of reversibility in the viscous flow. The final loss of reversibility requires the detailed analysis of the point in time where the system decides to keep or to leave a local structure. As a consequence, the viscosity is an independent process. This forces one to go over to the compliance description of eq. (\ref{jom}).

The present work shows that the Maxwell time marks the boundary between relaxation processes owing their long relaxation times to a high barrier (on the fast side) and relaxation processes owing their long relaxation time to the large number of relaxation processes which they require (the Maxwell time is itself the first example of this second class). At the same time, it is the boundary between back-and-forth motion and irreversible flow (the paper by Zwanzig \cite{zwanzig} aims at a correct treatment of exactly this crossover).

There is a sizable number of recent numerical attempts to find out what happens microscopically on the viscosity time scale \cite{sastry,furukawa,egami,oettinger}. On the accessible relaxation time scale of nanoseconds, one finds a dynamic correlation length via a four-point correlation function \cite{berthier}, which seems to diverge with diverging relaxation time according to the Adam-Gibbs recipe \cite{adam}. The present work identifies this correlation length with the core diameter of an Eshelby rearrangement \cite{eshelby}. Studies of the shear stress correlation \cite{furukawa,egami} show the strongly nonlocal character of the viscous process, in good agreement with the central postulate of the present paper. It is not quite clear whether the diverging correlation length dominates the viscosity \cite{furukawa} or whether the viscous flow requires even more space \cite{egami}, but in any case faster relaxations seem to occur on smaller length scales \cite{furukawa}. 

The advantage of the numerical work is that one can look at details of the atomic motion which are not accessible to experiment. The work of the swiss group \cite{oettinger} focuses on the correlation length for particles with a parallel displacement vector in deformation-induced transitions between inherent states. Consider a pure shear Eshelby rearrangement where the core expands in the $x$-direction and contracts along $y$. The displacements are four moving clouds of particles, each of them partly inside and partly outside of the core, two moving outside along $x$ and two moving inside along $y$, with a diameter similar to the one of the core itself. At the same time, there is a density excess in the $x$-clouds and a density diminution in the $y$-clouds, visible in the four-point correlation of the density. Thus one would expect the displacement correlation length \cite{oettinger} to be close to the dynamic density correlation one \cite{sastry}, as one indeed finds.

\section{Summary}

The present work describes an attempt to analyze the microscopic processes at and around the Maxwell time. The analysis bases on the asymmetry model \cite{asymm}, which assumes relaxational jumps from one unstrained ground state to many strongly distorted states of an inner core of about hundred atoms or molecules (a "cooperatively rearranging region" \cite{adam}).

The analysis relates the viscosity to the long-time cutoff of the retardation processes. The viscosity is ascribed to distorted states staying for such a long time that the surroundings adapt, making them new undistorted ground states.

In this picture, the Maxwell time $\eta/G_\infty$ ($\eta$ viscosity, $G_\infty$ infinite frequency shear modulus) is the borderline between reversible faster processes, with a relaxation time determined by a thermally activated jump over a barrier, and irreversible slower processes, which attain their long relaxation times by the necessity to jump over many barriers. The viscous flow itself belongs to this second category, which forces one to use a compliance description. "With compliance comes comprehension" - Don Plazek was right \cite{plazek-fin}.

\end{document}